# Three-dimensional modelling of processes in Electron Cyclotron Resonance Ion Source

V. Mironov[1], S. Bogomolov, A. Bondarchenko, A. Efremov, V. Loginov, D. Pugachev

*Joint Institute for Nuclear Research, Flerov Laboratory of Nuclear Reactions,*
*Dubna, Moscow Reg. 141980, Russia*
E-mail: vemironov@jinr.ru

Three-dimensional numerical model is developed and applied for studies of physical processes in Electron Cyclotron Resonance Ion Source. The model includes separate modules that simulate the electron and ion dynamics in the source plasma in an iterative way. The electron heating by microwaves is simulated by using results of modelling the microwave propagation in the plasma by the COMSOL Multiphysics® software. Extracted ion currents and other parameters of the source are obtained for different gas flows into the source. It is observed that the currents are strongly influenced by ion transport in transversal direction induced by the plasma potential gradients. Impact of some special techniques on the source performance is investigated. Magnetic field scaling is shown to reduce the ion losses during their movement toward the extraction aperture, as well as use of the aluminum chamber walls and mixing of the working gas with helium.

---

[1] The data that support the findings of this study are available from the corresponding author upon reasonable request.

In its essence, Electron Cyclotron Resonance Ion Source (ECRIS) [1] is an open non-axisymmetric plasma trap, in which electrons are resonantly heated by microwaves up to energies of ~(1-100) keV. Relatively cold (~1 eV) ions are produced in sequential ionizing collisions with electrons and reach high charge states. Those ions that leave the trap along the source axis are accelerated and used in a variety of applications. The extracted ion currents can be as high as a few mA for some ion species.

The currents are affected by many parameters, such as a configuration of the magnetic field, ad-mixing of other gases to the main working gas, use of multiple-frequency microwaves, the negatively biased electrode at the injection side of the source, and others [2]. For the optimized source, currents are increasing when adding more gas into the source and then saturate at some level. The same saturation is observed for increases in the injected microwave power. It is commonly conjectured that the saturation in the source performance occurs whenever the plasma density starts to be compared to the critical density for the microwaves with a given frequency. The efficiency of microwave coupling to the plasma is supposed to be affected by the high plasma density, thus limiting the source output.

The experimental observation is that the ion currents can be substantially boosted in the sources that use the microwaves with higher frequency; to the great extent, this frequency scaling defined the routes for the source development during the past decades [3]. To operate the sources with the higher microwave frequency, magnetic fields should be increased such as to provide the optimal ECR conditions in the plasma: the empirical magnetic field scaling laws require that the hexapole magnetic field at the source chamber walls ($B_{rad}$) should be at least twice larger of the resonant magnetic field ($B_{ECR}$), $B_{rad} \approx 2B_{ECR}$, the field at the extraction $B_{ext} \approx B_{rad}$, and the field at injection $B_{inj} \approx (3-4)B_{ECR}$. Thus, with doubling the microwave frequency, the source magnetic fields also should be doubled to keep the mirror ratios the same. The increase of the field suppresses the charged particle diffusion across the magnetic field lines, resulting in the better plasma confinement. It is questionable what is more important for the frequency scaling effect - the increasing of the magnetic field or the better microwave coupling.

To understand the mechanism of source performance saturation with the increased gas flow (plasma density), we perform numerical simulations of processes in ECRIS by using the NAM-ECRIS (<u>N</u>umerical <u>A</u>dvanced <u>M</u>odel of <u>E</u>lectron <u>C</u>yclotron <u>R</u>esonance <u>I</u>on <u>S</u>ource) group of codes. The model is extensively described elsewhere [4]. The major modifications in the current version are made in the parts concerning calculations of the electron dynamics. Now, to simulate the electron heating by microwaves we use COMSOL Multiphysics software® [5], which calculates the spatial and phase distributions for electromagnetic waves in the source chamber filled with the ECR plasma.

The main features of ECRIS operation are reproduced with the code in a self-consistent manner, without adjustment of any free parameters. From calculations, it can be deduced that the extracted ion currents are mainly limited by transport of ions across the magnetic field lines due to the formation of the plasma potential peak on the source axis. The transport rate is increased when adding more gas into the source due to increased plasma potential; a decrease in the potential is seen with adding more mobile ions such as helium. Also, by increasing the magnetic field, more ions can be extracted due to the better ion perpendicular confinement.

The paper is organized in the following way: first, we describe the general layout of the code, with emphasis on starting conditions and on communication between different modules. Then, details of the calculations with COMSOL Multiphysics® software are given. Electron dynamics is discussed in the next section, and afterward the charge state distributions of the extracted ion currents are presented for different gas flows into the source.

## The general layout of the model

The NAM-ECRIS is a combination of two 3D particle-in-cell modules that separately and sequentially trace dynamics of electrons and ions. The electron module requires as an input: 1) the microwave electric field spatial and phase distributions for simulations of the electron heating; 2) coordinates and energies of electrons that are created in ionizing collisions with heavy particles to initiate the calculations and to return electrons into the computational domain after their losses; 3) spatial distributions of electron and the charge-resolved ion densities to simulate electron-electron, electron-ion collisions, as well as electron energy losses due to ionization/excitation events. Electron densities are obtained in the previous run of the module, the ion densities are imported from the ion module, and the microwave field information is obtained by the COMSOL Multiphysics® model.

The electron module's outputs are the electron density spatial distribution, the electron energy spectra along the source axis and at the peripheral parts, and the globally defined electron life time. The electron densities are imported into the COMSOL Multiphysics® model, which recalculates the microwave fields. The ion module uses the output of the electron module to simulate the ion dynamics. The ion production and heating rates are defined by the input electron densities. Also, these densities are used by the 3D Poisson solver of the module to obtain the plasma electric potential distribution and to calculate the ion motion in the internal electric fields inside the plasma. These fields are directed such as to minimize a difference between the local space charge densities of ions and electrons. The statistical weight of computation particles is adjusted during the ion module operation to ensure that the ion life is close to the electron life time.

The obtained data on the ion densities and on the new-born electrons are then imported in the electron module and the iteration process continues until the converged solution is obtained. The iteration process is illustrated in Fig.1, where transversal (at the minimum of the magnetic field close to the source chamber center, (a)) and longitudinal (along the source axis,(b)) distributions of electron density are shown for a sequence of iterations. Typically, it takes ~5-10 iterations to obtain the converged solution.

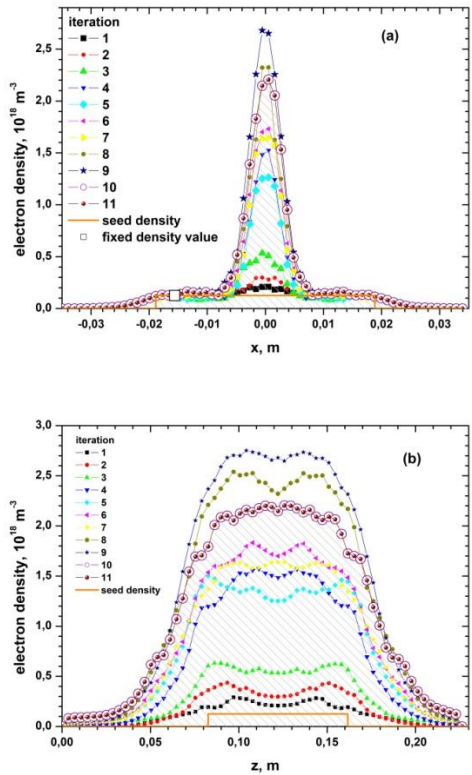

Fig.1 Transversal (z=11.5 cm, (a)) and longitudinal (along the z-axis,(b)) profiles of the electron density for different iterations.

The iteration cycle begins with setting the plasma density equal to a certain value inside the ECR volume and to zero outside the volume. The initial distributions for the seed electron density of $1.25 \times 10^{17}$ m$^{-3}$ are shown in Fig.1 as orange lines. Even after the very first iteration, electron spatial distribution starts to be strongly non-uniform, with a maximum along the source axis and relatively dilute halo. The maximum along the axis is due to the reduced electron losses caused by the retarding electrical fields close to the biased electrode and to the source extraction aperture. The module does not give the absolute values of the electron density, but the relative abundance of the numerical particles in the given computational mesh cell – we do not use a statistical weight of numerical electrons in the electron module. The relative spatial distribution of the electrons should be renormalized to get the absolute electron densities. For this, we request that the electron density at a rather arbitrarily selected point inside the plasma halo (close to the local maximum of the density at the source center) is always equal to the fixed value. The point is shown in Fig.1 as the open square. This reference density value characterizes the specific simulation set. Each iteration requests calculations for ~1 ms of the physical time, and starts with initial electron energies and positions prepared in the ion module. The solution is considered as converged if variations in the electron density and electron energies are less than 10%. The hashed curve in Fig.1 represents the converged solution of the electron module.

Simulations are done for the reference electron densities of 1.25, 2.5, 5 and $10 \times 10^{17}$ m$^{-3}$. For each converged solution, gas flow into the source is calculated. It is found in calculations that the larger is the reference electron density, the larger is the gas flow: the flows are 2.0, 3.2, 7.3 and 15.3 pmA (particle-mA) for the densities given above. Finally, correspondence between the injected gas flow and the global plasma parameters (including the extracted ion currents) is found in each converged iteration cycle.

We use the magnetic field distribution and the source chamber dimensions for the DECRIS-PM source [6]. The microwave frequency is set to 14.51 GHz according to the source operational parameters. The source chamber is 23 cm in length and 7 cm in diameter. The magnetic field of the hexapole is 1.1 T at the source walls; the solenoidal magnetic field on the source axis is 1.34 T at the microwave injection side and 1.1 T at the extraction side of the source. The extraction aperture is 1 cm in diameter. The biased electrode is installed at the injection, with the diameter of 3 cm and the negative voltage regulated in the range from 0 to -500 V depending on the source tunings. The source body is biased up to 25 kV in respect to ground for the ion extraction. In simulations, we use the value of -250 V for the biased electrode and set the extraction voltage to 20 kV. All calculations are done for the fixed injected microwave power of 500 W; responses of the source outputs to variations in the injected power are left for further studies.

## Modelling of the microwave propagation in the source plasma

For simulations of the electron dynamics in the source, we should take into account interaction of the particles with the electromagnetic waves injected into the source. The waves propagate in the magnetized anisotropic plasma located inside the metal chamber. We follow the approach of [7] for simulations of the wave propagation in ECRIS plasma by using the RF module of COMSOL Multiphysics® software [8]. Stationary frequency-domain 3D model is developed with anisotropic dielectric tensor calculated in the cold-plasma approximation.

The relative permittivity "cold" tensor $\ddot{\varepsilon}_{bz}$ can be written as [9]

$$\ddot{\varepsilon}_\updownarrow \equiv \begin{bmatrix} \varepsilon_{xx} & \varepsilon_{xy} & \varepsilon_{xz} \\ \varepsilon_{yx} & \varepsilon_{yy} & \varepsilon_{yz} \\ \varepsilon_{zx} & \varepsilon_{zy} & \varepsilon_{zz} \end{bmatrix} = \begin{bmatrix} S & -iD & 0 \\ iD & S & 0 \\ 0 & 0 & P \end{bmatrix} \quad (1)$$

for the magnetic field directed along the z-axis.
The components S, D and P are defined as

$$S \equiv \frac{1}{2}(R+L); D \equiv \frac{1}{2}(R-L); P = 1 - \tilde{n}_e$$

$$R = 1 - \frac{\tilde{n}_e}{1-\tilde{B}}; L = 1 - \frac{\tilde{n}_e}{1+\tilde{B}}$$

Here, $i$ is the imaginary unit, $\tilde{n}_e$ and $\tilde{B}$ are the electron density normalized to the critical density for the given microwave frequency $f_{RF}$ and module of the magnetic field normalized to the ECR resonance value,

$$n_{crit} = \left[\frac{f_{RF}}{8.98 \times 10^3}\right]^2 [cm^{-3}]; B_{ECR} = \frac{f_{RF}}{28 \times 10^9}[T]$$

The R-term in the tensor is infinitely large at the ECR resonance, which causes divergence of the COMSOL Multiphysics® solver. Therefore, we follow recommendations of the COMSOL Multiphysics® developers [8] and use the R-term in the form:

$$R \rightarrow R = 1 - \frac{\tilde{n}_e}{1 - i\Delta - \tilde{B}}$$

where $\Delta$ is positive Doppler broadening factor. We note here the negative sign in front of the broadening factor, which comes from the COMSOL Multiphysics® convention that negative imaginary terms in the dispersion relation correspond to the wave absorption. The $\Delta$-factor is a free parameter in the model. We observed that variations of the factor in the range from 0.005 to 0.05 did not change the solutions substantially; the factor is fixed to 0.02 throughout the calculations. More details on the $\Delta$-factor selection are given below. We do not include the collisional dissipation terms into the tensor because the collisional frequencies of the electron-ion and electron-electron scattering in typical ECRIS plasmas are in the range of $(10^4-10^5)$ Hz, much less than the cyclotron frequencies of around $\sim 10^{10}$ Hz.

For the arbitrary orientation of the magnetic field, we rewrite the tensor (1) by applying the following transformation [10]:

$$\tilde{\varepsilon} = U^T \cdot \tilde{\varepsilon}_\updownarrow \cdot U$$

where U is the transformation matrix in the form

$$U = \begin{bmatrix} \cos\theta\cos\varphi & \cos\theta\sin\varphi & -\sin\theta \\ -\sin\varphi & \cos\varphi & 0 \\ \sin\theta\cos\varphi & \sin\theta\sin\varphi & \cos\theta \end{bmatrix}$$

and $U^T$ is the transpose matrix of U.

The angles $\varphi$ and $\theta$ in the matrix are defined through the following relations:

$$\tan\varphi = \frac{B_y}{B_x}; \tan\theta = \frac{\sqrt{B_x^2 + B_y^2}}{B_z}$$

where $B_{x,y,z}$ are components of the local magnetic field vector.

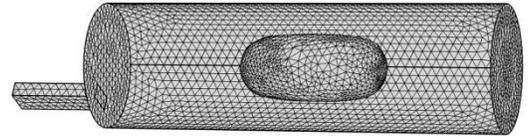

Fig.2 Geometry of the COMSOL Multiphysics® model.

Calculations are performed in a cylindrical chamber with the sizes that correspond to the DECRIS-PM geometry (Fig.2). Microwaves with 14.51 GHz frequency are injected through WR62 waveguide operating in $TE_{10}$ mode. The chamber walls are considered to be stainless steel, the waveguide material is copper, with the corresponding electrical conductivities. To have an opportunity to fine-tune the computational mesh, the computational domain is divided in two parts separated by the ECR surface. This surface is prepared by an external meshing program and is imported into the COMSOL Multiphysics® model. In Fig.2, the zone is shown as an ellipsoid in the chamber center.

During the calculations, the mesh size outside the zone is set to $\lambda/4$, where $\lambda$ is the wavelength of the microwaves in vacuum; the mesh size inside the zone is $\lambda/5$. The mesh is defined based on a compromise between the solution accuracy and execution time/available computer memory.

The magnetic field parameters and electron density spatial distribution are the inputs for the COMSOL Multiphysics® model. Normalized components of the magnetic field and electron density are prepared as 3D arrays with dimensions of $(100 \times 100 \times 100)$ in x-, y- and z-directions. The arrays are imported by the COMSOL Multiphysics® model to define the corresponding interpolation functions. The user-defined relative permittivity anisotropic tensor is then calculated with these functions and used to obtain the solution.

The model output includes the 3D arrays with the dimensions of (100×100×100) in x-, y- and z-directions, which contain (x,y,z) components of the microwave electric field amplitude, as well as the corresponding phases of the wave. The arrays are imported into the electron module of NAM-ECRIS for further calculations.

As the first step in the COMSOL Multiphysics® calculations, we check our model by obtaining the electromagnetic field distributions with injecting either the right-hand (RH) or left-hand (LH) circularly polarized wave into the source chamber. For this, we modify the injection scheme by launching the waves coaxially through a circular port with a diameter of 7 cm, same as the chamber diameter. For clarity, we set the radial wall conductivity to zero, thus minimizing the wave reflections from these parts of the chamber; the "extraction" side wall is fully reflecting. The electric field amplitude distribution for the case of no plasma present in the chamber is shown in Fig.3(a). The wave enters into the chamber from the left, is reflected from the opposite side and leaves the chamber through the injection port. The standing wave pattern in the Fig.3(a) is due to interference between the injected and reflected waves. The distribution does not depend on the wave polarization.

Next, the model is run with the electron density of $2.5 \times 10^{17}$ m$^{-3}$ inside the ECR volume, and with no plasma outside (the normalized density profile is shown in Fig.3(b)). The distributions of the electric field amplitude of the wave in x-z (y=0) plane of the chamber are shown in Fig.3(c)-(h) for the RH (Fig.3(c),(e),(g)) and LH (Fig.3(d),(f),(h)) injected waves for different $\Delta$-factors of 0.001, 0.02 and 0.05. The RH-wave is strongly reflected and absorbed at the ECR surface. For the LH-wave, the plasma is almost transparent for the chosen under-dense electron density.

For a small $\Delta$-factor of 0.001, dubious peaks of the electric field amplitude are formed beyond the ECR surface indicating numerical instability of the COMSOL Multiphysics® solver. An increase in the $\Delta$-factor dumps these irregularities; solutions are close each other for $\Delta$ in the range from 0.02 to 0.05, with small decrease in the field amplitudes for larger $\Delta$ due to increased wave absorption in the plasma regions near the resonant magnetic field surface.

Now we return to the description of calculations with the default geometry of the COMSOL Multiphysics® model (Fig.2), in which injection of the linearly polarized microwaves is done through the rectangular WR62 waveguide. Spatial distribution of the electric field amplitude for the empty chamber is shown in Fig.4(a). There, a typical standing wave pattern is seen with the electric field amplitudes reaching the level of around $1.5 \times 10^5$ V/m.

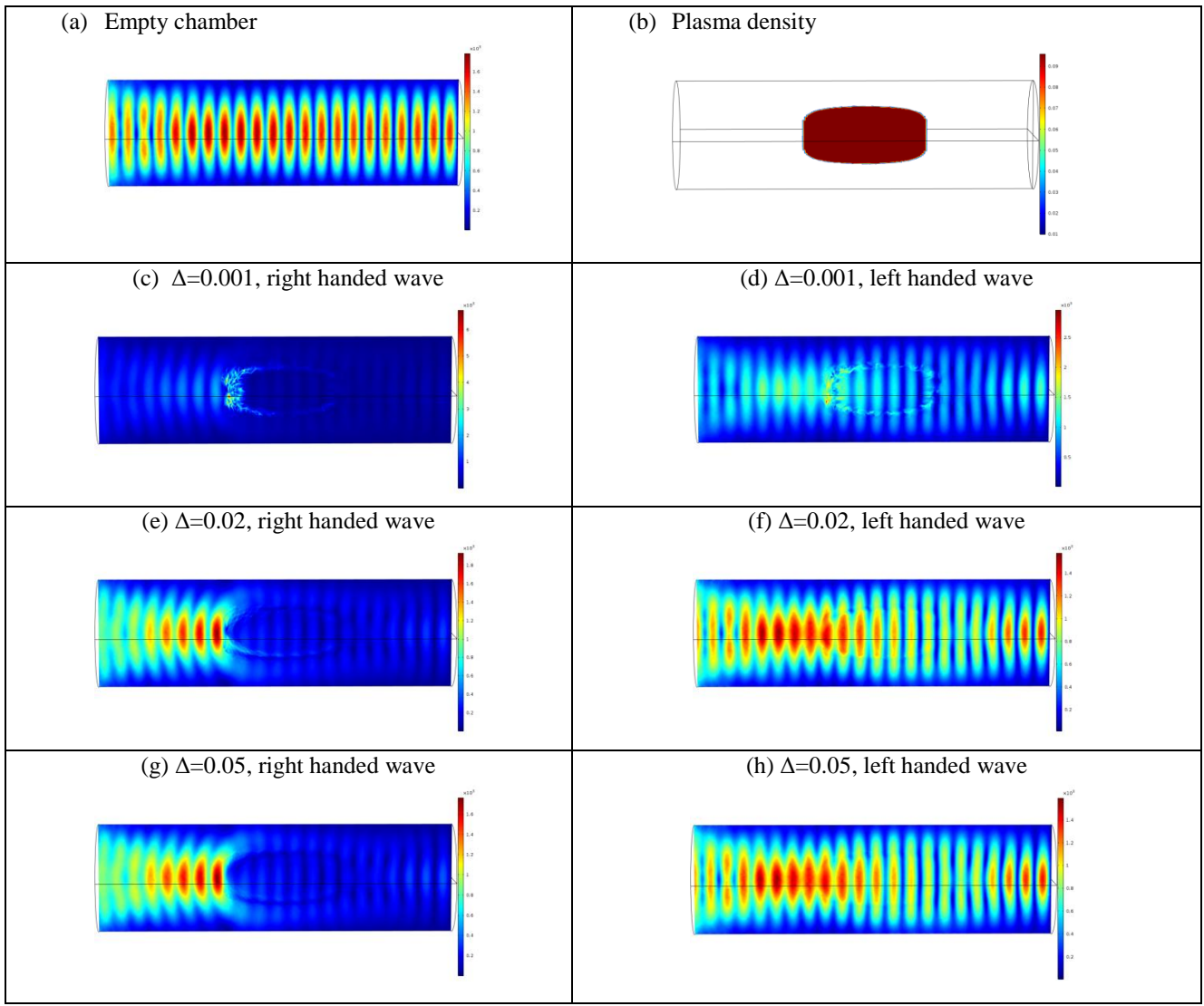

Fig.3 The COMSOL Multiphysics® calculations with injection of right- and left-handed polarized waves: (a) the microwave electric field distribution in the empty chamber; (b) plasma density distribution with uniform density of $2.5\times10^{17}$ m$^{-3}$ inside the ECR volume; (c),(e),(g) the the microwave electric field distribution for the right-handed wave with the Doppler factor of 0.001, 0.02 and 0.05; (d),(f),(h) the microwave electric field distribution for the left-handed wave with the Doppler factor of 0.001, 0.02 and 0.05.

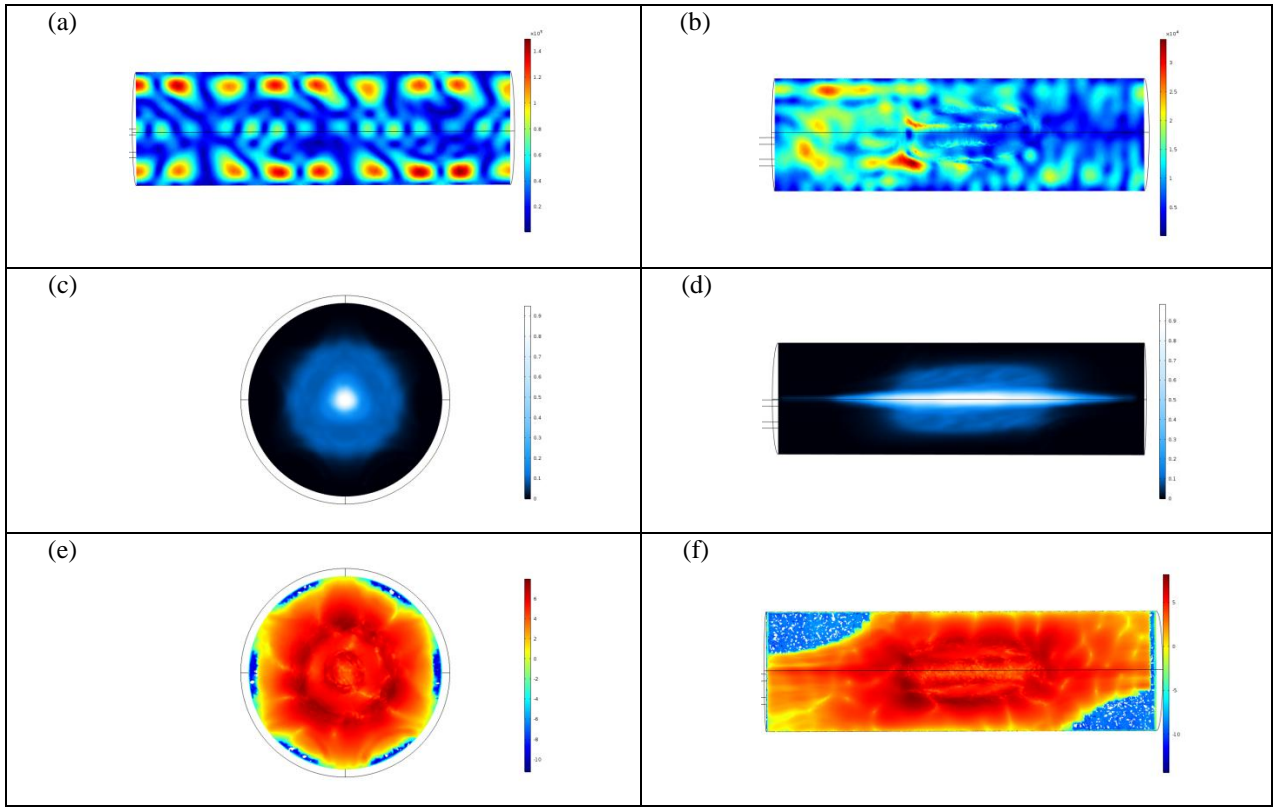

Fig.4 Longitudinal distribution of the microwave electric field amplitude for empty chamber (a); the same distribution for the converged plasma profile with the reference density of $2.5\times10^{17}$ m$^{-3}$ (b); transversal distribution of the normalized electron density, z=11.5 cm (c); longitudinal distribution of the normalized electron density (d); microwave power losses in the transversal direction, logarithmic scale (e); microwave power losses in the longitudinal plane, logarithmic scale (f).

Results of calculations with the converged plasma density distribution (for the reference plasma density of $2.5\times10^{17}$ m$^{-3}$) imported from the electron module are presented in Fig.4(b)-(f). In Fig.4(b), electric field amplitude distribution is shown, with the maximal values of around $3.5\times10^{4}$ V/m, strongly reduced in comparison to the empty chamber. Standing waves are still present in the picture; fields inside the ECR volume are substantially lower than in the peripheral parts. Transversal and longitudinal distributions of the electron density are shown in Fig.4(c)(d), while in Fig.4(e)(f) the absorbed power density profiles are shown in logarithmic scale. For this specific case the microwave power absorbed by the plasma is 471 W, power deposition on the chamber walls is 3 W, and the rest of the injected 500 W of microwave power is reflected back. For all investigated reference plasma densities, these power deposition values remain close each other. Also, we see that microwave electric field amplitude distribution is not changed strongly with variations in the reference density/gas flow into the source. Partially, this is due to saturation of the electron density in the central axial parts of the plasma, which will be discussed later. The plasma halo density is varying following the changes in the reference electron density, but it remains relatively small in the investigated range to cause significant microwave refraction and absorption.

Axial distributions of the electric field amplitude are shown in Fig.5 for the investigated set of the reference electron densities. The ECR positions are labeled with the dashed lines. The field amplitudes close to the resonance are at the level of ~$10^{4}$ V/m and decrease by a factor of 2 inside the ECR zone. Standing wave patterns are seen both inside and outside the zone regions.

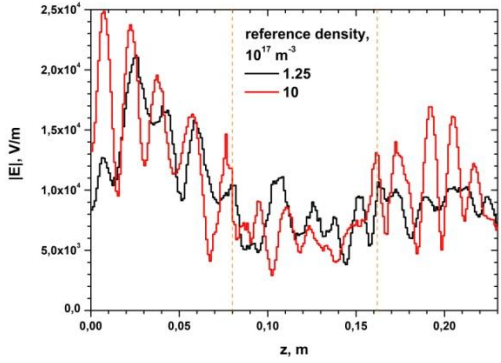

Fig.5 Axial distributions of the microwave electric field amplitude for the reference electron densities of 1.25 (black) and $10\times10^{17}$ m$^{-3}$ (red). The ECR zone positions are shown as dashed lines.

We note here that the presented distributions of the microwave electric field amplitude do not directly reflect the electron heating rate at given positions because the polarization information is missed there. The distributions are useful only in combination with the maps of the corresponding wave phases.

## Modelling of the electron dynamics

The electron module traces the movement of electrons in the source magnetic field and in the microwave electric field by using the relativistic Boris mover. The magnetic field is calculated by using the POISSON/SUPERFISH group of codes [11] for the solenoidal component; the hexapole component is calculated analytically as the field of the Halbach array in the hard-edge approximation. The microwave amplitudes $E_{0xyz}$ and the phases $\varphi_{xyz}$ are imported from the COMSOL Multiphysics® solution.

The electric field components are calculated each time step by linear interpolation between the values at the mesh nodes; the values vary in time as

$$E_{xyz} = E_{0xyz}\cos(2\pi f_{RF} t + \varphi_{xyz}),$$

where $f_{RF}$ is the microwave frequency of 14.51 GHz.

Influence of the internal electric fields is neglected because the electron energies are much higher than the typical values of the potential inside plasma; the electron dynamics in the plasma sheath regions is taken into account by elastically reflecting the particles from the computational domain walls if their energies along the magnetic field lines are less than 20 keV at the extraction aperture with the diameter of 1 cm, 250 eV at the biased electrode with the diameter of 2 cm, and 25 eV at other boundaries corresponding to the estimated plasma potential drop in the sheath.

Electron-electron and electron-ion collisions are taken into account by using the Spitzer's collision rates. Electron densities for collisions with electrons are taken from the previous iteration run of the electron module. The ion density maps are imported from the preceding iteration run of the ion module. Inelastic collisions (excitation and ionization) are simulated by using the cross-sections from ALADDIN database [12] and GKLV cross-sections for ionization [13]. The collisions are performed each $10^3$ time-steps to accelerate calculations and to minimize the round-off errors. We use a relatively small number of the computational particles ($10^3$) and the time step of $10^{-11}$ sec. This allows tracing the electrons for ~1 ms of the physical time with execution time of around 24 hours per iteration. At that, numerical noise in the calculated electron density distribution is minimized by using the under-relaxation and the cloud-in-cell charge deposition techniques.

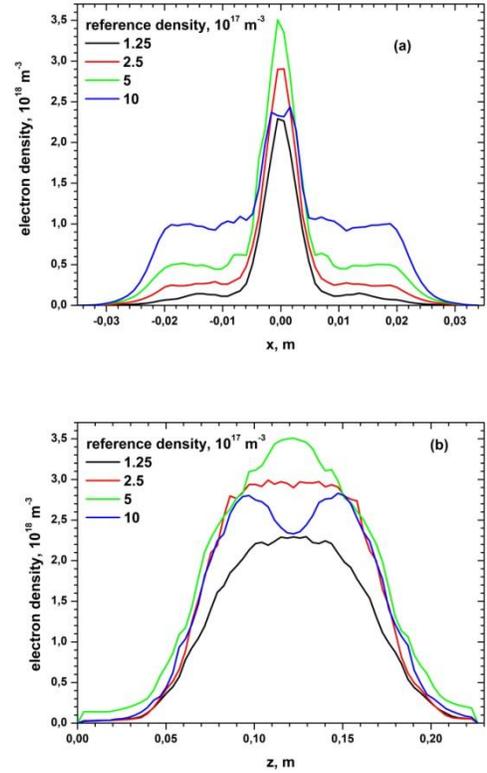

Fig.6 Transversal (a) and longitudinal (b) distributions of the electron density for the reference densities of 1.25, 2.5, 5 and $10\times10^{17}$ m$^{-3}$.

The main results of the electron simulations are shown in Fig.6(a)(b) for the spatial distributions of the electron density for the investigated set of the reference electron densities and in Fig.7 for the electron energy distribution functions (eedf) for the reference electron density of $2.5\times10^{17}$ m$^{-3}$.

The transversal distribution of the electron density (Fig.6(a)) includes the central plasma core, which is formed along the magnetic field lines terminated by the retarding potentials at the extraction aperture and the biased electrode. Density of this core plasma is above

$2\times10^{18}$ m$^{-3}$ for all reference densities, being maximized at the level of $3.5\times10^{18}$ m$^{-3}$ for the reference density of $5\times10^{17}$ m$^{-3}$. Ratio between the densities close to the axis and in the halo is decreasing with increased reference density, i.e. with the increased gas flow into the source. The halo plasma is localized inside the ECR volume in the transversal directions.

The same localization is seen in the longitudinal direction along the source axis (Fig.6(b)). The density profile is almost uniform inside the ECR volume, with formation of the small local density maxima close to the ECR surface for the largest gas flow. These maxima are much less pronounced in comparison to what had been reported in our previous publication [4], where the electron dynamics was calculated in assumption that electrons do not interact with microwaves inside the ECR volume.

The electron energy distributions in Fig.7 are shown separately for the core plasma (red) and for the halo (black). The distributions can be fitted with a sum of two exponentially decaying curves with the decay indexes of 7.5 keV and 75 keV in both parts of the plasma. Contributions of the 7.5-keV exponent are 0.15 for the core and 0.1 for the halo. The energy distributions are weakly dependent on the reference density/gas flow into the source.

The cold electron component is present in the spectra with the eedf best fitted by the 5-eV Maxwell-Boltzmann distribution. The cold electrons contribute ~5% into the total number of electrons in the plasma.

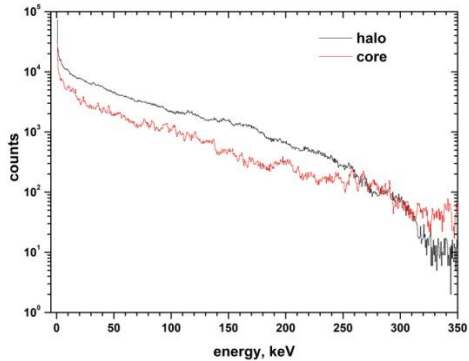

Fig.7. The electron energy distributions for the core plasma (red) and for the halo plasma (black) at the reference electron density of $2.5\times10^{17}$ m$^{-3}$.

For the halo electrons, a knee is seen in the spectrum at the energy of 250 keV (the relativistic factor $\gamma=1.5$). The origin of this knee is the increased loss rate for those electrons that enter into the cyclotron resonance with the microwaves near the source walls [4] due to the relativistic shift in the cyclotron frequency. The result of these increased losses is a hump in the energy spectrum of the lost electrons at 250 keV. Generally, the calculated energy distribution of the lost electrons strongly deviates from eedf of electrons that stay in the plasma: in the energy interval from 50 to 200 keV, the lost electron eedf is best described as the decaying exponent with the index of 35 keV, i.e. factor of two less than for the confined electrons.

The spatial dependence of the electron energy distributions is only partially taken into account in our model. The electron density profiles are calculated separately for electrons with energies below 100 eV (cold electrons), from 100 eV to 1 keV (warm electrons) and for the energies above 1 keV (hot electrons). From the density profiles, we see deviations from the eedf shown in Fig.7 in the regions beyond the ECR volume; electron population is noticeably colder there. When simulating the electron and ion dynamics in the plasma, these variations can be neglected because the electron density in these regions is small.

The electron energy distribution close to the extraction aperture is of special interest. Energies of electrons there directly influence the ion extraction by defining the plasma meniscus shape. The eedf for this plasma region is shown in Fig.8. The cold 5-eV component is seen that contains ~10% of all particles; the rest of particles forms the exponential tail with the index of $(220\pm15)$ eV (black curve). We remind that the biased electrode voltage is set to (-250) V in our calculations. The electron energies are directly affected by the biased electrode voltage – if we calculate the same distribution for the biased electrode voltage of (-50) V, then the tail index decreases to $(42\pm5)$ eV. The electrons close to the ion extraction region are much more energetic than it is typically assumed in the models [14].

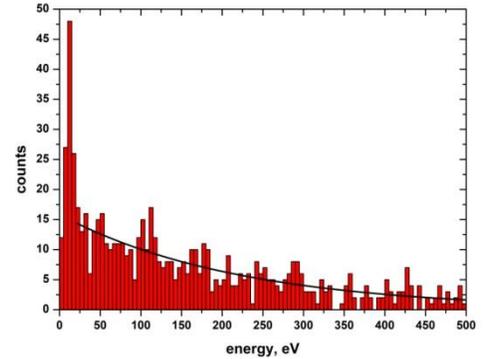

Fig.8. The energy distribution of electrons close to the extraction aperture. The fit with decaying exponent with the decay index of 220 eV is shown as the black line.

Electron life time changes significantly with changing the electron reference density. For the lowest density of $1.25\times10^{17}$ m$^{-3}$ the life time is calculated to be 0.32 ms, and then it decreases to 0.26, 0.17 and 0.12 ms for the densities of 2.5, 5 and $10\times10^{17}$ m$^{-3}$ respectively. The electrons in the core plasma are relatively well confined by the electrical potentials at the extraction aperture and the biased electrode. The halo electrons are confined weaker in the source magnetic mirror trap;

with increasing the reference density (gas flow), the core plasma density is not changing strongly, while the halo plasma density increases and the total losses are becoming faster.

Most of the electron losses are caused by interaction of electrons with the microwaves that pushes electrons into the loss-cone whenever they are passing through the electron cyclotron resonance. Electron losses due to other processes are much smaller: if we set the microwave electric field amplitudes to zero during the calculations, the electron loss rate is decreased by a factor of approximately 3 almost immediately after switching the microwave heating off. This abrupt decrease of the electron losses was experimentally observed elsewhere [15].

## Modelling of the ion dynamics

The ion module of NAM-ECRIS simulates the ion dynamics by taking into account the ion movement in the source magnetic field and in the electrostatic plasma potential, elastic and ionizing ion-ion and ion-electron collisions and energy losses of the heavy particles after collisions with the walls. Ion-ion collisions are simulated by using the Takizuka-Abe algorithm of pairing the colliding particles in a computational cell [16] and the Nanbu method for determining the scattering angles after collisions [17]. Charge-exchange collisions between ions and neutral particles are simulated by using the Langevin rates, with the kinetic energy releases calculated from the reaction energetics. The ionization rates are calculated with the GKLV cross-sections [13] using the electron energy distributions from the electron module. The excitation-autoionization rates are calculated by using the data from [18].

We assume that particles are leaving the source after going through the extraction aperture of 1-cm diameter. Also, we take into account that the injection plug of DECRIS-PM is perforated for better gas pumping, with the pumping channels covering ~50% of the plug surface for those regions that are not shielded by the biased electrode. Therefore, the particles that hit the injection side at positions beyond the biased electrode are considered either to be lost from the system with probability of 50% or being reflected from the wall. The lost particles are injected back into the chamber at the gas injection position to keep the number of the computational particles constant.

For the thermal accommodation at the walls we use the experimental data of Trott et.al. [19]. For the stainless steel surfaces, the thermal accommodation coefficients were measured as 95% for argon and 46% for helium. Whenever ions hit the walls, their primary energies are calculated as $25 \times Q$ [eV] for all surfaces except the biased electrode, where the energies are $250 \times Q$ [eV] -- the plasma potential is estimated to be of 25 V and the biased electrode voltage is set to 250 V according to the typical operational conditions of the source. After neutralization on the walls, reflected particles retain a significant fraction of their primary energy, which is important for the ion dynamics in the plasma both from the point of view of ion heating and of the neutral particles penetration into the dense parts of the plasma.

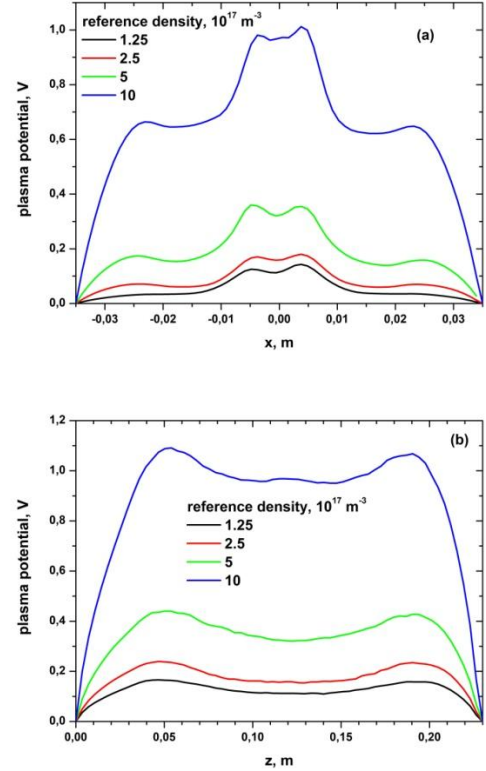

Fig.9 Transversal (a) and longitudinal (b) distributions of the plasma potential for the reference densities of 1.25, 2.5, 5 and $10 \times 10^{17}$ m$^{-3}$.

Ions are moving in internal electric fields that buildup to maintain the plasma quasi-neutrality. The fields are calculated with 3D Poisson solver, details of the calculations are given in [4]. The plasma potential is determined with subtraction of the plasma potential drop of ~25 V in plasma sheath. The plasma potential profiles are shown in Fig.9 in the transversal direction at the plasma center (a) and along the source axis (b) for the investigated range of the gas flows into the source (reference electron densities).

The potential is positive everywhere in the plasma with maximum values in the plasma core close to the source axis. In transversal direction, large gradients of the potential are seen at the core boundaries and in the regions between the plasma halo and the radial walls. These transversal electric fields strongly influence the ion movement, push ions away from the source axis and decrease the extracted ion currents. In the longitudinal direction, relatively small plasma potential dip is formed inside the ECR volume, which decreases the ion losses from there. The dip value is around 0.1 V and depends on the plasma parameters. Beyond the ECR volume, electric field accelerates ions toward the walls following the electron density drop in these regions.

The plasma potential is expected to be larger for plasma with the heavy lowly charged and cold ion component compared to plasma with more mobile ions. Also, the higher is the ion production rate, the larger electric fields are required to push ions out of the plasma for equilibration of electron and ion life times and for keeping the electron and ion densities close to each other. This explains the observed increase in the plasma potential with increased gas flow in Fig.9. For the low gas flows of 2.0 and 3.2 pmA, the potential in the plasma center is ~0.15 V and it increases up to ~1 V for the largest gas flow of 15.3 pmA.

The main calculated parameters of the plasma for different gas flows are listed in Table 1. The global (averaged over the full ion population) ion temperature is increasing with the gas flow from 0.2 eV to 0.5 eV for $Ar^{8+}$ ions, being strongly influenced by the ion acceleration in the plasma electric fields. Dependence of the global ion temperatures on their charge state is close to linear. Temperature of neutral argon atoms in the source chamber is increasing from 0.036 to 0.056 eV in the investigated range of gas flows, indicating increase of the ion fluxes to the chamber walls. The mean charge state of argon ions in the plasma is only slightly increasing with the gas flow from 2.38 to 2.54: faster ionization of ions in the denser plasma is counterbalanced by decrease in the ion life time.

Table 1. Main calculated parameters of the source plasma

| | | | | |
|---|---|---|---|---|
| Reference electron density, $10^{17}$ m$^{-3}$ | 1.25 | 2.5 | 5 | 10 |
| Gas flow, particle-mA | 2.0 | 3.2 | 7.3 | 15.3 |
| Electron/ion life time, ms | 0.32 | 0.26 | 0.17 | 0.12 |
| Plasma potential in the centre, V | 0.11 | 0.16 | 0.32 | 0.96 |
| Plasma potential dip, V | 0.05 | 0.07 | 0.12 | 0.13 |
| Temperature of argon atoms, eV | 0.036 | 0.045 | 0.053 | 0.056 |
| Temperature of $Ar^{8+}$ ions, eV | 0.19 | 0.28 | 0.42 | 0.5 |
| Mean charge state of ions in the plasma | 2.38 | 2.45 | 2.52 | 2.54 |
| Extraction efficiency for $Ar^{8+}$ ions (extracted/total flux), % | 22 | 17 | 12 | 6.5 |
| Total extracted ion current, mA | 1.8 | 2.5 | 4.6 | 7.1 |
| Mean charge state of the extracted ions | 3.83 | 3.78 | 3.34 | 2.8 |

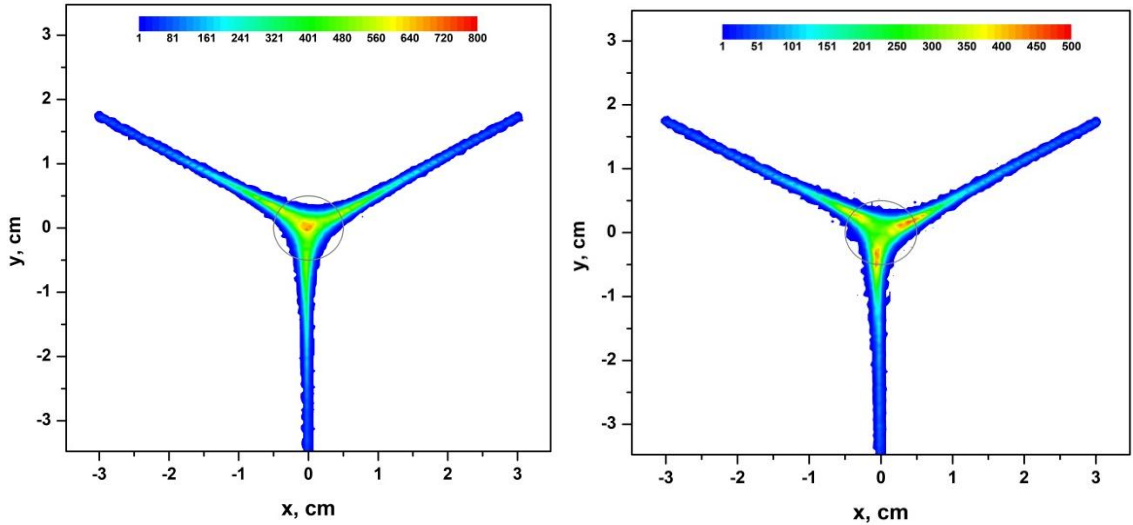

Fig.10 Positions of the $Ar^{8+}$ ions hitting the extraction electrode for the reference densities of 2.5 and $10\times10^{17}$ m$^{-3}$.

Influence of the transversal electric fields on the ion transport in the plasma is illustrated in Fig.10, where the $Ar^{8+}$ ion positions after hitting the extraction electrode are shown for the gas flows of 3.2 pmA and 15.3 pmA. The extraction aperture is shown in the Figure as the grey circle. For the relatively small gas flow and the plasma potential, the profile is well centered on the source axis and the transversal dimensions of the ion distribution inside the extraction aperture are small. For the large gas flow and plasma potential, the profile is strongly distorted, rotated in the direction of the E×B drift and becomes hollow. Substantial displacement of ions from the source axis along the plasma star arms results in decrease of the extracted current.

For the low gas flow and relatively small plasma potential, the fraction of $Ar^{8+}$ ions that pass through the extraction aperture is ~22% of the total ion flux out of the plasma. This value is decreasing by factor of ~4 to 6.5% for the large gas flow – ion extraction efficiency becomes low in these conditions. This drop in the extraction efficiency is caused not only by transport in the plasma potential electric field, but also by the fact that most of ions at large gas flows are produced in the plasma halo, not in the plasma core (Fig.6) and are flowing along the magnetic field lines to the radial walls of the source chamber.

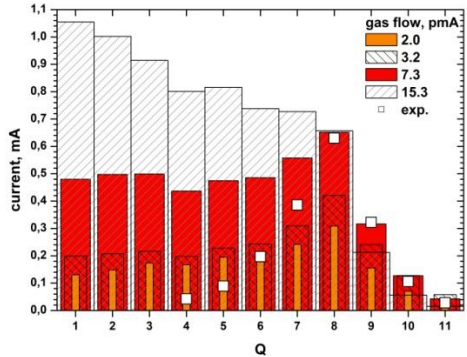

Fig.11 Calculated charge state distributions of extracted ion currents for the argon gas flows of 2.0, 3.2, 7.3 and 15.3 pmA (the reference densities of 1.25, 2.5, 5 and $10\times10^{17}$ m$^{-3}$). The typical experimental data are shown as the open squares.

The charge state distributions of the extracted argon ions are shown in Fig.11 for the investigated range of the gas flow variations. The total extracted ion current increases with the gas flow from 1.8 mA to 7.1 mA, while the mean charge state is decreasing from 3.8 to 2.8. The currents for the highest charge states saturate and then decrease at the largest gas flow. Currents of $Ar^{8+}$ ions do not exceed ~0.7 mA close to the maximal observed values for the DECRIS-PM source [6]. The experimental points are shown in the Fig.11 as open squares; the simulated distribution at the gas flow of 7.3 pmA is close to the experimental data for the charge states 8+ and higher. For the lower charge states, the measured currents are substantially less than the calculated values, probably because of the ion beam losses in the beam transport line, which was tuned for transmission of the highly charged ions when measuring the currents.

We see that the internal electric fields greatly influence the ion dynamics in ECRIS plasma. It is instructive to investigate the influence on the source output of some techniques widely used to boost the ECRIS performance. The results are shown in Fig.12. As the reference points, currents of argon ions are shown there (open squares), which were calculated for the electron density of $5\times10^{17}$ m$^{-3}$, the same data as presented with the red columns in Fig.11.

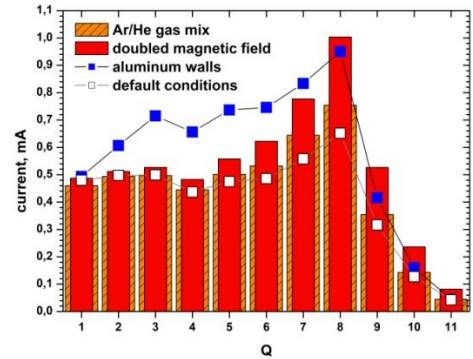

Fig.12 Calculated charge state distributions for the reference density of $5\times10^{17}$ m$^{-3}$ for the gas-mixed plasma (orange), for the doubled magnetic field (red), for the aluminum chamber walls (blue squares) and for the plasma in default computational conditions (open squares).

First, we simulate the source performance after scaling the magnetic fields of the source by factor of two, leaving all other parameters (electron density distribution, electron life time etc.) untouched. The higher magnetic field slows down the ion transport across the magnetic field lines, which results in the substantial increase in the ion extraction efficiency from 12 to 18%. Extracted currents of $Ar^{8+}$ are increased from 0.65 mA to 1.0 mA, while currents of lowly charged ions are not changing substantially. Charge state distribution of the extracted ion currents with doubled magnetic field is shown in Fig.12 as red columns. The plasma potential is not changed with the magnetic field scaling, being mostly defined by the ion transport along the magnetic field lines, ion heating and production rates that remain the same as in the reference conditions.

The presented data indicate that the frequency-scaling effect can be at least partially explained by reduction of the transversal transport of ions in the increased magnetic field.

Output of ECRIS is known to be affected by the chamber wall conditions and material. In particular, it is observed that the aluminum chambers allow increasing the extracted ion currents [20]. The wall-coating effect is explained by higher secondary electron emission from the walls that increases the electron life time and density in the plasma. Another explanation [20] is connected with the experimentally observed decrease of the plasma potential in the sources with aluminum chambers, which presumably leads to the better plasma stability due to reduced sputtering of the walls by ions.

We notice that the thermal accommodation coefficient in collisions of argon with aluminum is much lower than for collisions with the stainless steel. In [21], it is measured that the coefficient is around 0.81 for aluminum compared to 0.95 for the stainless steel. Gas temperature in the aluminum chamber should be relatively high due to the slower absorption of the particle excess energy in collisions with the walls. This process results in higher energies of ions inside the plasma, in decrease of the plasma potential that expels the particles out of the plasma and consequently in reduced transversal ion transport.

In Fig.12, the ion charge state distribution is shown that is obtained using the thermal accommodation coefficient of argon for the aluminum walls (blue squares). The plasma parameters are kept the same as for the reference calculations with the electron density of $5\times10^{17}$ m$^{-3}$. Currents of ions are substantially boosted compared to the default conditions with the stainless steel walls. At the same time, plasma potential in the center is decreased from 0.3 to 0.2 V, while the extraction efficiency for Ar$^{8+}$ ions is increased up to 20%. The potential dip along the source axis is 0.175 V, substantially larger than 0.12 V for the default conditions. The argon gas temperature is increased up to 0.085 eV and temperature of Ar$^{8+}$ ions reaches the level of 0.78 eV. The gas flow into the source increases up to 8.8 pmA compared to 7.3 pmA for the default conditions. Gain in the extracted ion currents is most pronounced at the relatively low charge states.

The experimentally observed gas-mixing effect is the increase of extracted currents of the highly charged heavy ions when adding a light gas to the plasma [2]. In [22], the effect was explained as a sequence of the heavy ion life time increase after injection of diatomic molecular gases such as oxygen and nitrogen, which ionization and dissociation lead to creation of highly energetic atoms and singly charged ions. Presence of such fast particles heats the whole ion population and increases the plasma potential dip to equilibrate the electron and ion losses out of the plasma. For the highly charged ions, increase in the ion temperature is smaller than increase in the potential dip, which results in longer ion confinement.

In calculations of [22], no boost in the argon ion currents is observed when using helium as the mixing gas. In practice, however, such boost is seen and often used for optimization of ECRIS operation for materials with moderate masses, for which mixing with oxygen or nitrogen is not effective. To further investigate the gas-mixing effect in the present version of the NAM-ECRIS model, we calculate the source output in a mix of argon and helium at the plasma with the reference electron density of $5\times10^{17}$ m$^{-3}$ (orange columns in Fig.12). It is found that the Ar$^{8+}$ ion current is increasing with adding helium as the mixing gas and reaches 0.75 mA for (3:1) ratio of total numbers of Ar and He computational particles in the chamber. This corresponds to the argon gas flux of 7.0 pmA and the helium gas flow of 5.6 pmA, while the helium extracted currents are 0.32 and 0.13 mA for He$^{1+}$ and He$^{2+}$ ions respectively. The plasma potential at the center is decreased to 0.24 V, the plasma potential dip along the source axis is 0.11 V, which is the same as in the default conditions. The ion temperature for Ar$^{8+}$ is 0.36 eV in the gas-mixed plasma compared to 0.42 eV for the default conditions. The helium temperatures are 0.045, 0.26 and 0.4 eV for He$^0$, He$^{1+}$ and He$^{2+}$ respectively. Extraction efficiency for Ar$^{8+}$ ions is increasing in the gas-mixed plasma to 16% compared to 12% for the non-mixed argon plasma, indicating slower transversal transport of ions due to the decreased plasma potential. We see no substantial changes of the argon ion life time in the plasma mixed with helium.

## Conclusions

After incorporating the COMSOL Multiphysics® calculations of the microwave electric fields inside the ECRIS plasma, the NAM-ECRIS model becomes essentially to be free of adjustable parameters tuned to numerically reproduce the source performance. This increase reliability of the model calculations and allows studying responses of a source to variations of input parameters. In particular, we examined the DECRIS-PM behavior with changing the gas flow into the source. It is found that the maximal extracted ion currents are limited by transversal transport of ions caused by the plasma potential gradients. The calculated ion currents are close to the experimental values. We connect the beneficial impact of the magnetic field scaling, wall effect and gas-mixing effect in ECRIS to the plasma potential reduction and to reduction of the transversal transport of ions. These observations are useful for optimization of existing sources and for finding the ways to design the new sources.

## Acknowledgement

This work was supported by the Russian Foundation for Basic Research under grant No. 20-52-53026/20.